\documentclass[aps,pre,twocolumn, superscriptaddress,showpacs,amsmath,amssymb]{revtex4}

\usepackage{graphicx}
\usepackage{dcolumn}
\usepackage{bm}

\usepackage{epstopdf}
\usepackage{color}

\usepackage{bm}
\usepackage[normalem]{ulem}

\usepackage[space]{grffile}

\newcommand{\bv}{\mathbf{v}}

\newcommand{\bbr}{\mathbf{r}}

\newcommand{\tri}{\triangle}

\newcommand{\beq}{\begin{equation}}
\newcommand{\eeq}{\end{equation}}
\newcommand{\beqn}{\begin{eqnarray}}
\newcommand{\eeqn}{\end{eqnarray}}
\newcommand{\pp}{\partial}
\newcommand{\dd}{{\rm d}}

\newcommand{\eq}{Eq.\ }

\newcommand{\cF}{{\cal F}}

\newcommand{\Sec}{Sect.~}

\newcommand{\eqs}{Eqs }
\newcommand{\fig}{Fig.\ }

\newcommand{\cO}{{\cal O}}

\newcommand{\la}{\langle}

\newcommand{\ra}{\rangle}

\newcommand{\vnab}{{\bf \nabla}}

\newcommand{\cred}{}

\begin{document}
	
	
	\title{
	Interface stability, interface fluctuations, and the Gibbs-Thomson relation in motility-induced phase separations}
	\author{Chiu Fan Lee}\email{c.lee@imperial.ac.uk}
\affiliation{Department of Bioengineering, Imperial College London, South Kensington Campus, London SW7 2AZ, U.K.}

	\date{\today}
	
	\begin{abstract}
		
	Minimal models of self-propelled particles with short-range volume exclusion interactions have been shown to exhibit signatures of phase separation. Here I show that  the observed interfacial stability and fluctuations in motility-induced phase separations (MIPS) can be explained by modeling the microscopic dynamics of the active particles in the interfacial region. In addition, I demonstrate the validity of the Gibbs-Thomson relation in MIPS, which provides a functional relationship between the size of a condensed drop and its surrounding vapor concentration. As a result, the coarsening dynamics of MIPS at vanishing supersaturation follows the classic Lifshitz-Slyozov scaling law at the late stage.
	\end{abstract}
	
	\maketitle

\section{Introduction}

Phase separation is a ubiquitous phenomenon in nature and is  manifested by the partitioning of the system into compartments with distinct properties, such as the different particle densities in the two co-existing phases in the case of liquid-vapour phase separation. Phase separation under equilibrium dynamics is a well investigated physical phenomenon  
\cite{domb_b83, bray_r02}. Recently, signatures of phase separation have been reported in non-equilibrium systems { consisting of} active particles
\cite{tailleur_prl08,cates_pnas10,peruani_prl11,clewett_prl12,farrell_prl12,fily_prl12,schwarz_pnas12,cates_epl13,redner_prl13,bialke_epl13,buttinoni_prl13,stenhammar_prl13, wittkowski_natcomm14,grosberg_pre15, stenhammar_prl15,bialke_prl15,sear_a15}. It is therefore a natural question to ask to what degree we can extend our knowledge of equilibrium phase separation to the phase separation phenomenon observed in active systems. 
{
In the case of minimal models of active particles with simple volume exclusion interactions, the phenomenon of motility-induced phase separations (MIPS) has  received considerable interest \cite{tailleur_prl08,peruani_prl11,fily_prl12,schwarz_pnas12,cates_epl13,redner_prl13,bialke_epl13,buttinoni_prl13,stenhammar_prl13, wittkowski_natcomm14,grosberg_pre15, stenhammar_prl15,bialke_prl15}.
}
   In particular, the idea of an effective surface tension in motility-induced phase separations (MIPS) has been advocated \cite{wittkowski_natcomm14,bialke_prl15}.   
At the {\cred gas-liquid} interface in thermal equilibrium, surface tension results from the pulling of molecules at the interface due to their attractive interactions \cite{marchand_ajp11}. In a system of active particles with purely repulsive interactions, it is unclear how such ``pulling'' can occur as the particles can only push.   To probe what happens at the interface, I study here the microscopic dynamics of the active particles in the interfacial region by a combination of simulation and analytical methods. Specifically, using mean-field type arguments, I will demonstrate how { pressure} balance is achieved between the condensed phase and the dilute (vapor) phase, and how the Gibbs-Thomson relation arises in a system where a circular condensed drop co-exists with the vapour phase. Furthermore, by incorporating the stochastic nature of particle dynamics, I will explain the scaling  between the interfacial width and the system size recently observed in MIPS \cite{bialke_prl15}.

\subsection{Motility-induced phase separation}  
I will first focus on a minimal 
model system that exhibits MIPS in two dimensions (2D)---A collection of self-propelled particles with excluded area interactions that undergo rotational fluctuations. Specifically, the dynamical equations are
\beqn
\frac{\dd \bbr_i}{\dd t} &=& -\frac{1}{\eta}\sum_{j\neq i}\vnab_{\bbr_i} U(|\bbr_i-\bbr_j|) +  \frac{f_a}{\eta}\bv_i
\\
\frac{\dd \theta_i}{\dd t} &=& \sqrt{2D_r} g_i(t)
\eeqn
where $i$ is an integral index enumerating the particles in the system, $\bv_i \equiv \cos \theta_i \hat{x}+ \sin \theta_i \hat{y}$ with the angle  $\theta_i$ (with respect to the $x$-axis) being the orientation of the $i$-particle, $g_i(t)$ is a noise term with { Gaussian probability distribution} with zero mean and unit variance, $D_r$ sets the magnitude of the rotational fluctuations, $U(.)$ corresponds to the potential function for short-ranged area exclusion interactions, $\eta$ is the drag coefficient and $f_a$ is the constant active force that drives the particles in the system. In particular,
$u\equiv f_a/\eta$ is the constant speed of a particle when it is not within the area exclusion zone of another particle. 
Previous numerical work has indicated that phase separation in this minimal system occurs  as $u$  increases, but the actual form of $U$ is unimportant \cite{fily_prl12,redner_prl13,stenhammar_prl13}.  For instance, $U$ could be of the form of a Weeks-Chandler-Andersen potential \cite{weeks_jcp71}:
\beq
\label{eq:wca}
U(r) = 
\left\{
\begin{array}{ll}
	A\left[ \left( \frac{a}{r}\right)^{12}-2\left(\frac{a}{r}\right)^{6})+1\right] \ , & {\rm if} \ r<a
	\\
	0 \ , & {\rm otherwise}
	\ .
\end{array}
\right.
\eeq
This will be the particular form of potential function employed in this work. Also, the time and length units will be set by having $a=1$ and $D_r=3$.
Note that I will focus exclusively on the non-equilibrium dynamics of the system and so 
translational Brownian motion is ignored.

\section{Flat interface}
\label{sec:flat}
\subsection{Point particles}
\begin{figure*}
	\centering
	\includegraphics[scale=.71]{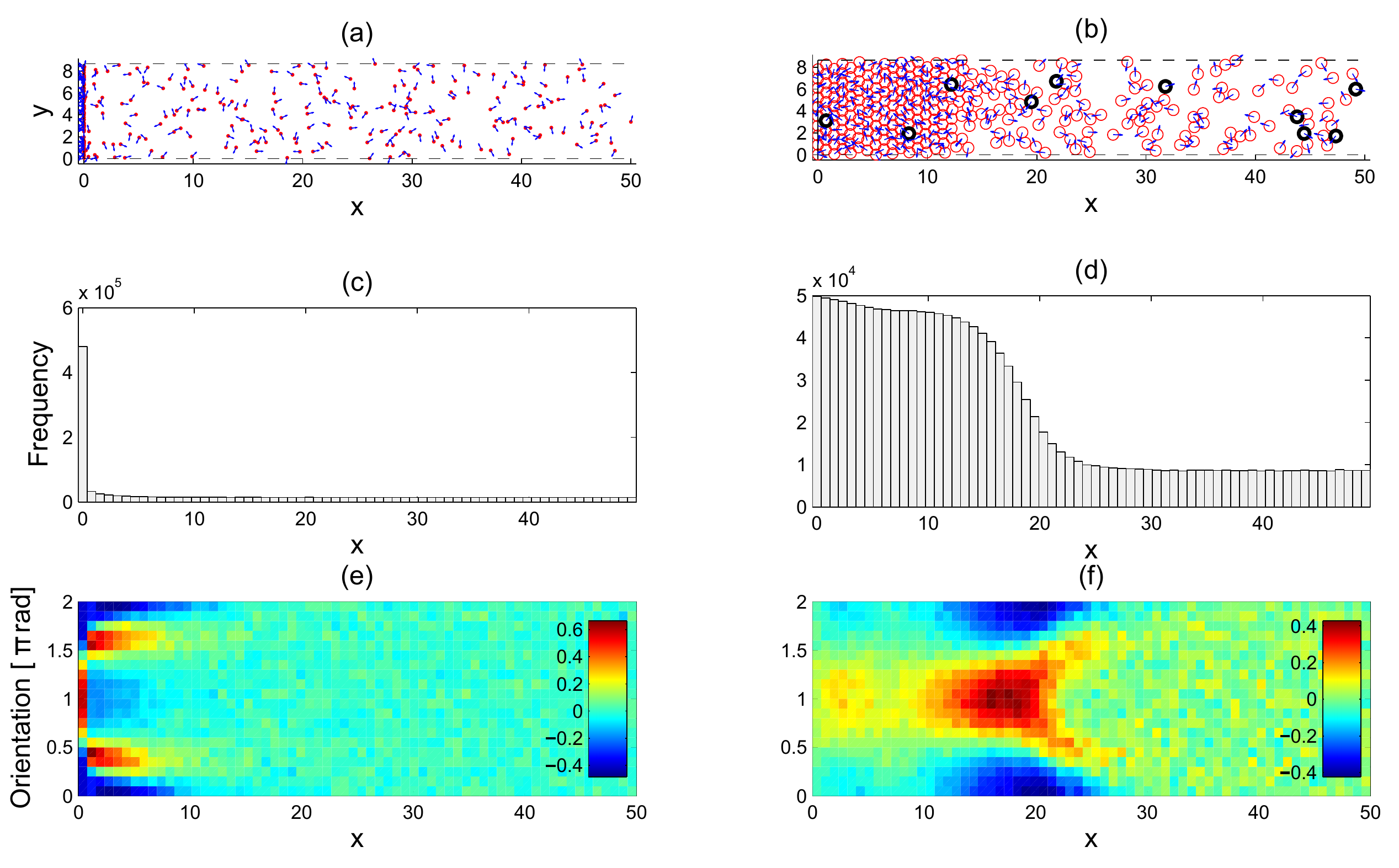}
	\caption{Steady-state configurations of active particles confined by a force absorbing wall on the left: point particles (left column) and repulsive particles (right column). a) \& b): A snapshot of the system at the end of the simulations with the orientations depicted by the blue arrows. The red circles depicted in (b) are of diameter $a=1$. The wall at $x=0$ is perfectly force absorbing  (see  \ref{sec:a1} for simulation details).  c) \& d): The histograms show the horizontal distributions of the particles. e \& f): The colourmaps show the deviation from the mean in the particles' orientations at different horizontal positions. The colour scale corresponds to the measure: $2(h_i(x) -\la h_i(x) \ra)/\max_i h_i(x)$ where $i$ is the row index and $h_i(x)$ is the frequency. The simulation parameters are: $f_a=100, \eta=1, a=1, D_r =3, A=25/6$.
	}  \label{fig:mainpic}
\end{figure*}

To understand the microscopic dynamics at the interface, it is instructive to first look at a system of point particles, i.e., the interaction potential $U$ is zero. Even this simple system distinguishes itself from equilibrium system in that aggregation will spontaneously happen in the proximity of a force-absorbing but frictionless wall (left column of \fig \ref{fig:mainpic}).
{\cred  In other words, the particles are free to slide and rotate at the wall, but they cannot penetrate the wall. }
 Further complex patterns are revealed when one looks at the particles' orientation distribution as well as the position distribution (\fig \ref{fig:mainpic}(e)). At the wall, most of the particles are left going, as indicated by the high concentration of orientation at around $\theta \simeq \pi$. 
This results from the fact that only left-going particles remain at the wall. Just outside the wall, the distribution is highly peaked at $\theta$ just below $\pi/2$ and just above $3\pi/2$, which reflects the particles' orientation after they move away from the wall. The orientation anisotropy decays as one moves away from the wall.

{ In this system, the pressure acting on the wall can be expressed as 
\beq
P_W= \left| f_a\int_{\pi/2}^{3\pi/2}\chi_W(\theta)  \cos \theta \dd \theta \right|
\ ,
\eeq
where the  orientation distribution function of the particles at the wall per unit length is denoted by $\chi_W(\theta)$.} { Note that since we are dealing with a 2D system, the unit of pressure is [force]/[length].}

To further analyse $\chi_W(\theta)$, one can perform dimensional analysis to conclude that
\beq
\chi_W(\theta) = \frac{f_a \rho_\infty}{\eta D_r} \cF({\theta})
\ ,
\eeq
where $\cF({\theta})$ is a function dependent only on $\theta$, and $\rho_\infty$  is the particle concentration far from the wall. 
To obtain the exact functional form of $\cF(\theta)$, one needs to solve a set of two coupled differential equations with mixed boundary conditions {\cred \cite{lee_njp13}}, whose solution consists of a series of Mathieu functions. Unfortunately, the expansion coefficients in the series are not analytically tractable and so an analytical expression is lacking. However, $\cF(\theta)$
can be readily estimated numerically (left column of \fig \ref{fig:mainpic}), which allows us to obtain the following:
\beq
\label{eq:fw_pt}
{ P_W} = \left| \frac{f_a^2 \rho_\infty}{\eta D_r} \int_{\pi/2}^{3\pi/2} \cF(\theta) \cos\theta \dd \theta \right| = \frac{f_a^2 \rho_\infty}{2\eta D_r}
\ .
\eeq
 The second expression is equivalent to  the {\it swim pressure} of a system of active particles in 2D \cite{takatori_prl14,yang_softmatter14,mallory_pre14,solon_natphys15}, which I have obtained numerically here.

\subsection{Repulsive particles}
Remarkably, much of what we have seen in the point particle case remains true when we add mutually repulsive interactions to the particles. When a force absorbing but frictionless wall constitutes the left boundary of a semi-infinite system, phase separation occurs where the condensed phase is located close to the wall (\fig \ref{fig:mainpic}(b)). Inside the condensed phase, the orientation is  isotropic (\fig \ref{fig:mainpic}(f)). 
The reason behind the isotropy is that the impeded motility of the particles render them staying put for a duration much longer than the orientation decoherence time $\simeq 1/D_r$. However, note that the particles' locations are not frozen in time as shown by the black particles in \fig \ref{fig:mainpic}(b), which were the first column of particles next to the wall at the beginning of the simulation.  As one moves further to the right, one first encounter a layer of left-going particles (shown by the bright red patch centred at  $x\simeq 17$ in \fig \ref{fig:mainpic}(f)). This represents the accumulation of  particles with orientation highly centred at around $\theta \simeq \pi$, analogous to the accumulation of active point particles at the wall, except that the particles are now  spread over a range of $x$ positions due to volume exclusion interactions. Further rightwards, we encounter the 
pattern of two escape trajectories away from the interfacial region indicated by the two red-yellow branches emerging from the red patch.
 As one moves further away to the right, the orientation becomes isotropic again. From this discussion, it is clear that in the bulk of the condensed  and vapour phases, the corresponding orientation distributions are both isotropic, while in the interfacial region separating them there is a high level of orientation anisotropy. 

Let us now calculate of the force exerted on the wall by the  active force of these particles. Since in the condensed phase, the orientation is isotropic,  the { pressure felt by the wall due to these active forces is}
\beq
\label{eq:fw}
{ P_W^{(a)}} =  \left|\frac{a\rho_c f_a}{2\pi}\int_{\pi/2}^{3\pi/2} \cos \theta \dd \theta \right|= \frac{a\rho_c f_a}{\pi}
\ ,
\eeq
where $\rho_c$ is the concentration in the bulk condensed phase. Due to the orientation isotropy, the expression here 
{ scales like $f_a$ instead of $f_a^2$}
in the point particles system (\eq (\ref{eq:fw_pt})). Besides the active force contribution,  the wall will also feel additional forces arising from the repulsive interactions, which, we will see, constitute an important contribution  in achieving { pressure} balance in the interfacial region in MIPS (Sect.~\ref{sec:stability}).

\subsection{Locating the interface}

\begin{figure}
	\centering
	\includegraphics[scale=.52]{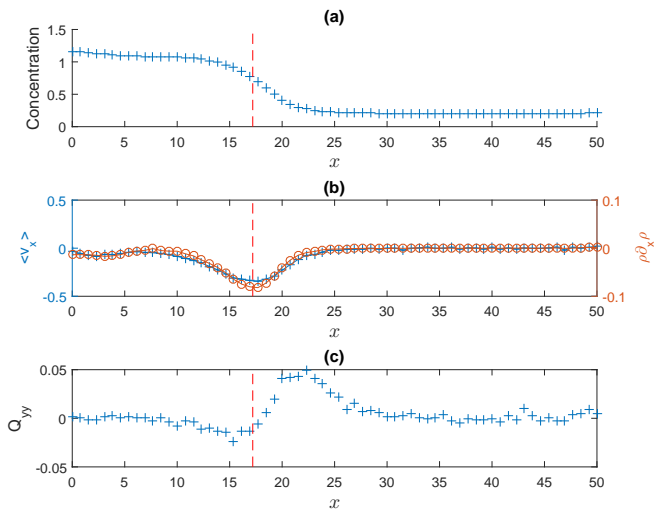}
	\caption{a) Particle concentration as a function of $x$. Same figure as in \fig \ref{fig:mainpic}(d). b) The mean horizontal component of the particles' orientations $\la v_x \ra$  vs.~$x$ ($+$ symbols, left $y$-axis), and $\rho\pp_x\rho$ vs.~$x$ ($\circ$ symbols, right axis).  c) The $yy$ component of the nematic order parameter $Q$ vs.~$x$, where $Q_{yy} \equiv -\la \cos(2\theta)/2 \ra$ so that high $Q_{yy}$ signifies that the orientations of the particles are { preferentially pointing up or down.}    
		Similar to equilibrium fluids, defining the location of a sharp interface is somewhat arbitrary \cite{hansen_b06}. The working definition proposed here is that the interface is set to be at the { pronounced minimum} of $|\la v_x \ra|$. 
	}  \label{fig:interfaces}
\end{figure}

The concentration variation across the two phases shown in \fig \ref{fig:interfaces}(a) is similar to typical equilibrium phase separation. What distinguishes MIPS is the high orientational anisotropy between the phases (\fig \ref{fig:interfaces}(b) \& (c)).  
As in  equilibrium fluids, the location of a sharp interface between the two phases can be defined somewhat arbitrarily  \cite{hansen_b06}. In our case, since the { pronounced minimum} of $\la v_x \ra \equiv \la \cos \theta \ra$ is easy to locate (indicated by the red broken line) and its location also marks the onset of the increase in  $Q_{yy} \equiv -\la \cos(2\theta)/2 \ra$ (the $yy$ component of the nematic order parameter $Q$) \cite{degennes_b95}, which signifies the escape of particles from the condensed phase,  it is a
convenient choice for the interface location. 
{ In other words, this convention implies that right outside the interface of the condensed phase, there is a layer of particles travelling preferentially along the interface, as indicated by the peak in $Q_{yy}$ (\fig \ref{fig:interfaces}(c)).
	The active forces of these escaped particles are potentially the cause of the emergence of a negative surface  tension according to its mechanical definition \cite{bialke_prl15}.
}
 The definition of the interface location has of course no physical significance, but this does provide a working definition useful for the sharp interface model discussed below.

\subsection{Interface stability}
\label{sec:stability}
\subsubsection{{ Pressure} balance at a sharp interface.~~}
 In this section we will see how { pressure} balance can be achieved in MIPS.
{ Note that the discussion in this section amounts to a simplified exposition of that in \cite{takatori_prl14,solon_prl15}.  Its presentation here is for the self-containedness of the paper and will help us understand the approximations used in later sections

I will start by discussing a sharp interface model. 
}
In this drastically simplified model, let us imagine that the phase separated system is partitioned by a sharp interface where in the vapour phase, the concentration is low enough that the system behaves like a system of active point particles, and in the condensed phase, the  orientation distribution is isotropic. One can imagine such a system by first rotating \fig \ref{fig:mainpic}(e) by $180^\circ$ and then collating it to \fig \ref{fig:mainpic}(f) on the left. As calculated before, 
The { pressure} exerted by the vapour phase on the sharp interface is $f_a^2 \rho_v/(2 \eta D_r)$ (\eq (\ref{eq:fw_pt})), this { pressure} is balanced by the { pressure} exerted by the condensed phase: $a\rho_c f_a/\pi + { P_r}$, where the first term comes from the active force (\eq  (\ref{eq:fw})) and the second denotes the { pressure} arisen from the repulsive force due to the area exclusion interactions. In other words, { pressure} balance is achieved if
\beq
\label{eq:forceb}
\frac{a\rho_c f_a}{\pi} + { P_r}=\frac{\rho_v f_a^2 }{2 \eta D_r}
\ .
\eeq
For the simulation parameters used in \fig \ref{fig:mainpic} (with the units set by $a=1$ and $D_r=3$), $a\rho_c f_a/\pi \simeq 30$, ${\cred P_r} \simeq 230$ and
$\rho_v f_a^2 /(2\eta D_r) \simeq 330$.  We thus see that  
in the simulated system, {\cred the L.H.S. and the R.H.S. of \eq (\ref{eq:forceb}) are of the same order of magnitude, indicating that the pressure balance condition is {\it qualitatively} satisfied}. In addition, we see that much of the  active force coming from the vapour phase is used to compress the condensed phase via the { pressure $P_r$}. 
Note that although \eq (\ref{eq:forceb}) provides a { pressure} balance condition for the system, it does not mean that any system satisfying this condition is stable as it may still be unstable against fluctuations. This is not dissimilar to equilibrium fluids where pressure balance {\it together} with chemical potential balance are needed to achieve phase stability.

The { pressure} balance condition  in \eq (\ref{eq:forceb}) already allows us to estimate crudely the minimal active force required for MIPS.
{\cred I assume for simplicity that at} the onset, i) the condensed phase is not very compressed and so we can ignore { $P_r$},\footnote{An anonymous referee has indicated that numerical simulation shows that $P_r$ is about a third of the other terms at the onset of MIPS. However,  
	\eq (\ref{eq:minf}) remains valid as $P_r$ does not depend on the P\'{e}clet number at the onset, as shown in the preceding paper} and ii) the concentration ratio between the condensed phase and the dilute phase (${\rho_c}/{\rho_v}$)  is of order 1, which let us to the minimal force requirement below for MIPS:
\beq
\label{eq:minf}
f_a^{min} \geq \frac{2a\eta D_r }{\pi}
\   .
\eeq
The above condition comes from the { pressure} balance condition at the interface alone. { Interestingly, \eq (\ref{eq:minf}) reproduces the same scaling as obtained by Redner {\it et al} {\it via} a different approximation \cite{redner_prl13}. The result also supports the notion that the P\'{e}clet number (Pe), usually defined as Pe $\propto   f_a/(a \eta {\cred D_r})$ in this context, is a key control parameter in MIPS \cite{redner_prl13,stenhammar_softmatter14}. 
}

From this sharp interface model, we can now see why the condensed phase with a high density of active and repulsive particles can remain stable against a backdrop of dilute concentration of active particles -- the active particles in the vapour phase impact an active { pressure} that scales as $f_a^2$ directed towards the normal of the interface, while the countering active { pressure} from the condensed phase scales as $f_a$.
The quadratic dependence in the  active force comes from the fact only particles pushing against the interface will remain on the interface   while particles with orientation away from the interface will leave. The escapes of these particles thus open up space for yet other particles that serve to push against the interface. Form this perspective, the low concentration in the vapour phase is paramount for the stability of MIPS, for otherwise the particles with orientations away from the interface may be unable to leave effectively.

\subsubsection{Force balance in an interface of finite width.~~}
Let us now go beyond the previous sharp interface picture and see what happens within the interfacial region from the view point of particle dynamics. 
Ignoring fluctuations, the stability of the interface  means that if  a particle happens to be lying at the interface will, on average, remain put. 
In our minimal model, a particle can only move due to two reason: i) its own active force driving it to move in the direction dictated by its orientation, ii) repulsive force that pushes it away from its neighbours if it is of less than unit distance away from them. While the second force is common in both active and passive (equilibrium) systems, the active force is unique to non-equilibrium systems. Consider now a particle  located at $x_0$ inside the interfacial region, i.e., where $\la v_x(x_0) \ra$ is varying (\fig \ref{fig:interfaces}). Since  $\la v_x(x_0) \ra<0$, the active force will on average drive this particle to the left. On top of this, there are repulsive forces coming from neighbouring particles on the right hand side $f_r(x_0+\tri x)$. For the particle to remain still, the sum of these forces has to be countered by the repulsive forces coming from the left. Therefore, 
\beq
\label{eq:dfr}
f_r(x_0-a/2) = f_a\la v_x(x_0) \ra+f_r(x_0+a/2)
\ .
\eeq
Since the repulsive forces come from the repulsive potential function $U$, let us replace the repulsive force by the pressure $P_p(x)$ (the subscript $p$ for passive) resulting from the corresponding system with the same particle configuration and interaction potentials, but with the active force omitted. Since $f_r(x)\simeq a P_p(x)$, \eq (\ref{eq:dfr}) leads to 
\beqn
f_a\la v_x(x_0)\ra &\simeq &a[ P_p(x_0+a/2)- P_p(x_0-a/2)]
\\
\label{eq:dPp}
&\simeq & a^2\frac{ \dd P_p(x_0)}{\dd x} 
\ .
\eeqn
In principle, $P_p(x)$ depends on the exact configuration of particles in the system, but if one adopts the simplifying assumption that the passive pressure depends solely on the particle concentration, one can then expand $P_p(x)$ with respect to the concentration $\rho(x)$:
\beq
P_p(x)=c_0+c_1 \rho(x)+c_2 \rho(x)^2 +\cO(\rho^3)
\ .
\eeq
For equilibrium fluids, this is of course the virial expansion where $c_0=0$ and $c_1=k_BT$ \cite{hansen_b06}. Since our system is fundamentally non-equilibrium (no translational Brownian motion, i.e., $k_BT=0$), there is no guarantee that the same would apply here. But let us assume that such an expansion is possible in our system, then since $P_p$ comes purely from the repulsive interactions between particles. We thus expect that $c_0=0=c_1$ because as the concentration goes to zero, there would not be any pairwise interactions. Therefore, the first non-trivial term in the expansion is $c_2\rho^2$.
{ 
	Note that $c_2>0$ since $P_p$ arises purely from the repulsive interactions. One could also incorporate active pressure into the analysis as, for example, done by Winkler {\it et al} 
	\cite{winkler_softmatter15}.
} 

{ Here, to  order $\cO(\rho^2)$,} \eq (\ref{eq:dPp}) then leads to
\beq
\label{eq:fa&Pr}
f_a\la v_x(x_0) \ra = 2a^2c_2 \rho(x) \frac{\dd \rho(x_0)}{\dd x}
\ .
\eeq
Remarkably, simulation result shown in \fig \ref{fig:interfaces}(b) indeed seems to vindicate \eq (\ref{eq:fa&Pr}).


\section{Circular interface}
\label{sec:curved}
We have seen in the previous section how { pressure} balance is achieved at a flat interface. However, previous 2D simulation studies have shown that similar to equilibrium phase separation, if the condensed phase in MIPS does not span the system size, the condensed phase  is circular. Here, we will see how the curvature of the interface  affects the particle dynamics at the interface, and its consequence in terms of the coarsening dynamics.  We will first study the emergence of the Gibbs-Thomson relation by dimensional analysis.

\subsection{Gibbs-Thomson relation: dimensional analysis}
\label{sec:gtrel}

In equilibrium phase separation, the Gibbs-Thomson (GT) relation dictates that the concentration $\phi_R$ right outside a droplet (of the condensed phase) of radius $R$ is
\beq
\label{eq:gt}
\phi_{R}=\phi_0 \left(1+\frac{\nu}{R}\right)
\ ,
\eeq
where  $\phi_0$ is the supersaturation concentration, i.e., the threshold concentration beyond which phase separation occurs, and $\nu=\frac{2\gamma v}{k_BT}$ is the capillary length with $\gamma$ being the surface tension and $v$ being the volume of the molecule. Since the concentration in the vapour phase outside a big drop is lower than that outside a small drop, a diffusive flux is set up that transfers material from the small droplet to the big droplet. This is the { Ostwald} ripening mechanism that dominates the phase separation kinetics at the late stage for systems with a small supersaturation \cite{lifshitz_jpcs61}.

I will now discuss why the GT relation would arise naturally in our active system. 
In the minimal MIPS system considered, the only parameters in the dynamical equations are the free roaming speed $u = f/\eta$, the rotational diffusion coefficient $D_r$,  and the length scale of the short range area exclusion interaction $a$. Denoting now the vapour density far from a drop of radius $R$ by $\rho_R^*$, and the density inside the drop (the condensed phase) by $\rho_c$, then by dimensional analysis we have
\beq
\frac{\rho_R^*}{\rho_c} = \cF\left(\frac{u}{D_ra},\frac{u}{D_rR} \right)
\ ,
\eeq
where $\cF$ is some unknown scaling function dependent on its two dimensionless arguments. If we now assume that $\cF$ is regular with respect to the second argument in the sense that a Taylor series expansion exists (around $u/D_rR=0$), then the ratio above can be re-expressed as 
\beq
\label{eq:rho_R0}
\frac{\rho_R^*}{\rho_c}  =H + K\frac{u}{D_rR} +\cO\left(\left(\frac{u}{D_rR}\right)^{2}\right) \ ,
\eeq
where $H$ and $K$ are now just dimensionless functions of the first argument $(u/D_r a)$, i.e., $R$-independent. In terms of $\rho_\infty^*$, \eq (\ref{eq:rho_R0}) can be re-written as
\beq
\label{eq:rho_R}
\rho^*_R =\rho^*_\infty \left(1+ \frac{\tilde{\nu}}{R}\right) +\cO\left(\left(\frac{u}{D_rR}\right)^{2}\right) \ ,
\eeq
where  $\tilde{\nu} \equiv \frac{Ku}{HD_r}$, which may be termed the effective capillary length. 
In the large $R$ limit, \eq (\ref{eq:rho_R}) becomes exactly the GT relation in \eq (\ref{eq:gt}). 
This analysis provides an intuitive reason why one would naturally expect  the GT  relation  to emerge as the drop radius grows in MIPS.

\begin{figure}
	\centering
	\includegraphics[scale=.33]{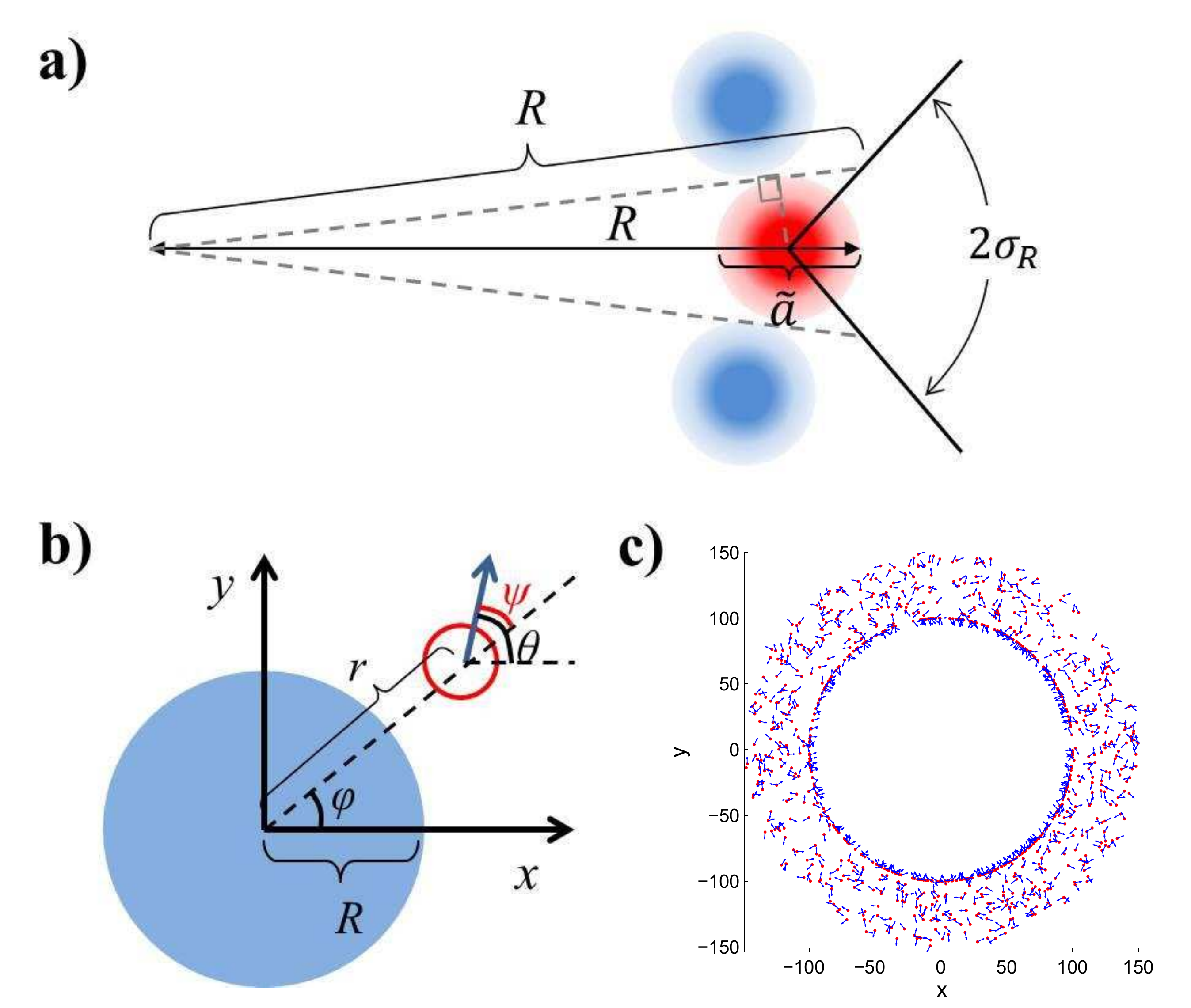}
	\caption{a) A schematic of the interfacial condition at a curved interface of curvature $R^{-1}$. The particle is assumed to occupy a zone of diameter $\tilde{a}$ and the particle can leave the droplet if its orientation is within the escape range of $2\sigma_R$ indicated. b) A schematic showing a droplet (blue) in the condensed phase of radius $R$  located at the origin co-existing with the dilute medium (vapour phase). An active particle (red circle) with orientation $\theta$ (blue arrow) is located at the position $(r, \varphi)$. The angle $\psi$ equals the difference between the orientation $\theta$ and the azimuthal coordinate $\varphi$.
		c) A snapshot of a simulated system with 1000 active point particles (red dots with orientations indicated by blue arrows) in an annular system with inner radius $R=100$ and outer radius $R+L_r=150$.
	}  \label{fig:Cinterface}
\end{figure}

\begin{figure}
	\includegraphics[scale=.45]{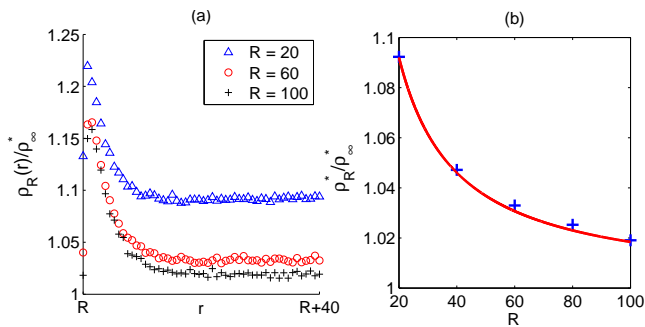}
	\caption{a) The variation of the vapour concentration $\rho_R(r)\equiv \frac{1}{r}\int \dd \psi \zeta_R(r, \psi)$ away from the interface of a droplet of sizes $R=20,60,100$.
		The concentration decays rapidly from the interface and reaches a stationary value $\rho_R^*$ a short distance away from the interface.  b) $\rho_R^*/\rho_\infty^*$ vs.~$R$. The curve decays to 1 like $R^{-1}$, which is consistent with the  Gibbs-Thomson relation (\eq (\ref{eq:rho_R})).  Blues crosses are the simulation results and the red curve depicts the function $1+1.84R^{-1}$. { Note that the constant $\rho_\infty^*$ is estimated numerically from the Gibbs-Thomson relation.}
		See \ref{sec:a2} for simulation details.
	}  \label{fig:res1}
\end{figure}

\subsection{Gibbs-Thomson relation: numerics}
\label{sec:gtrel2}

I will now test
\eq (\ref{eq:rho_R}) by simulating a coarse-grained model of MIPS.  Let us first consider what happens to an active particle in the condensed phase at the periphery that is curved (\fig \ref{fig:Cinterface}(a)). For such a particle, I assume that it occupies a zone of radius $\tilde{a}$ (shown in red) sandwiched by two zones occupied by two neighbouring particles (light blue). Note that since the concentration at the interface may not reach the level of optimal packing concentration ($\simeq 0.91$), $\tilde{a}$ should be greater than the particle's diameter $a$. Indeed, \fig \ref{fig:interfaces}(a) suggests that the concentration is around 0.72 at the interface, which indicates that $\tilde{a} \simeq 1.2$. To incorporate the effects of the neighbouring particles on the pink particle,  I  assume that as a result of the caging effect, the particle can only move out of the droplet if its orientation is within the $2\sigma$ range depicted. Based on the diagram shown in  \fig \ref{fig:Cinterface}, a simple trigonometric exercise leads to
\beq
\label{eq:theta}
\sigma_R = \frac{\pi}{4}+ \frac{3\tilde{a}}{2R} +\cO\left(R^{-2}\right)
\ .
\eeq
As expected, a lower curvature leads to a smaller escape range (smaller $\sigma_R$).

To analyse how the variation in the escape orientation range affects the phase separated system at the steady-state, I    
consider a system with one condensed drop of radius $R$ co-existing with the vapour phase (\fig \ref{fig:Cinterface}(b)). Let me denote the particle distribution function in the vapour phase by $p_R(r,\varphi, \theta)$ where the first two arguments correspond to the particles' locations  and the last argument to the particles' orientations. 
Due to rotational symmetry, one can eliminate one angular argument by introducing the variable  $\psi \equiv \theta -\varphi$ \cite{pototsky_epl12}, and study instead the reduced distribution function  $\zeta_R(r, \psi) \equiv \int_0^{2\pi}  p_R(r,\varphi, \varphi+\psi) r \cos \varphi \dd \varphi$. On the drop's periphery, the corresponding reduced distribution function is denoted by $\chi_R(\psi)$. Since the periphery is assumed to be infinitely thin, $\chi_R$ is only a function of $\psi$ and hence dimensionless. In addition, I assume that drops of all sizes  have the same interior concentration $\rho_c$,   $\chi_R(\psi)$  is thus related to the $\rho_c$ as follows: $\int \dd \psi \chi_R (\psi) \simeq 2\tilde{a}R \rho_c$.

To study the distribution functions, I assume again that the vapour concentration is low enough that pairwise repulsive interactions can be ignored, and simulate the dynamics of active point particles, { i.e., non-interacting active particles, in an annular geometry} of inner radius $R$ and outer radius $R+L_r$(\fig \ref{fig:Cinterface}(b)). As in the linear case, the particles' orientations are randomised if they reach the outer circular boundary, while if the $i$-th particle reaches the inner boundary, its positions will remain fixed until its orientation is within the escape range, i.e., until   
$\psi_i$ is between $-\sigma_R$ and $\sigma_R$.  Simulation results are shown in (\fig \ref{fig:res1}). Away from the interface,  it is observed that the concentration rapidly reaches a stationary value $\rho_R^* \equiv \frac{1}{r}\int \dd \psi \zeta_R (r,\psi)$ for, say, $r>R+20$ (\fig \ref{fig:res1}(a)). 
As expected from previous discussion, the vapor concentration $\rho_R^*$ goes down with $R$ since a flatter interface leads to a narrower escape range, which leads to a smaller outflux of particles from the condensed phase. 
\fig \ref{fig:res1}(b) shows that $\rho_R^*$ decays to $\rho_\infty^*$ (the vapour concentration as $R\rightarrow \infty$) like $R^{-1}$, which, as we have seen, is consistent the Gibbs-Thomson relation in equilibrium systems.

\begin{figure}
	\centering
	\includegraphics[scale=.5]{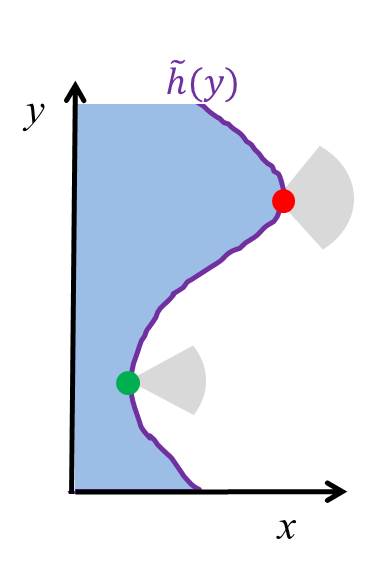}
	\caption{A schematic depicting a wavy interface where the condensed phase is depicted in blue. The location of the interface (purple) is given by the function $\tilde{h}(y)$. Due to the caging effect from neighbouring particles, the red particle at the interface will have a higher chance of escaping compared to the green particle because the escape orientation range (grey area) is bigger.
	}  \label{fig:Finterface}
\end{figure}

\section{Fluctuating interface}
\label{sec:fluctuations}
I have so far ignored fluctuations in the interfacial profile. In reality, the interface of course fluctuates, which is  already discernible from the spatially  constrained system shown in \fig \ref{fig:mainpic}(b). In particular, previous simulation result points to the scaling law \cite{bialke_prl15}:
\beq
w_L^2 \sim L
\ ,
\eeq
where $w_L$ is the steady state interfacial width:
\beq
w_{L_y}\equiv \frac{1}{L_y} \sqrt{\int_0^{L_y} (\tilde{h}(y)^2 - \bar{h}^2) \dd y}
\eeq
with $\bar{h}$ being the average position of the interface. Here, the symbol $\tilde{h}(y)$ denotes the location of the interface, i.e., the location of the peak of $\la v_x \ra$ (\fig \ref{fig:interfaces}). To understand the scaling observed, let us consider the effects of interface curvature on the particle exchange dynamics. Although the previous section focuses only on a circular interface, i.e., the curvature is positive, one can easily extends \eq (\ref{eq:theta}) to allow for concave interface as well (\fig \ref{fig:Finterface}). The physical motivation behind the formula is the same, a particle at a highly convex portion of the interface  will have a wider escape orientation range (red particle in \fig \ref{fig:Finterface}) than a particle at a highly concave interface  (green particle).

\begin{figure}
	\centering
	\includegraphics[scale=.59]{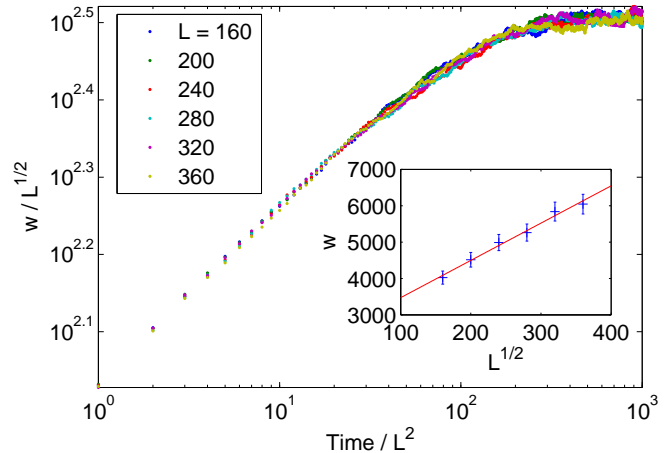}
	\caption{Interface fluctuations as measured by the interface width $w_L(t) \equiv \sum_{i=1}^L (\tilde{h}_i(t)-\bar{h}(t))^2$. The curve collapse of systems with difference linear dimension $L$ upon rescaling is as predicted by the EW model. {\it Inset plot:} The  interface width at the final time $w_L(t_f)$ shows a linear dependence with $\sqrt{L}$ where $L$ is the system size. Blues crosses are the simulation results and the red line is a guide for the eyes.  See Appendix \ref{sec:a3} for simulation details.
	}  \label{fig:ew}
\end{figure}

Since the fluctuations ultimately come from the fluctuating dynamics of particle exchange at the interface, we need to model the steady-state dynamics of $\tilde{h}$ stochastically. The simplest equation of motion (EOM) for the interface that incorporates both the effects of stochasticity and curvature-modified outflux  is
\beq
\label{eq:tildeh}
\frac{\pp \tilde{h}}{\pp t} =2\alpha b_1(y) - 2\alpha'(1+ \beta\kappa(y)) b_2(y)
\ ,
\eeq
where $\alpha$ denotes the rate of particles coming into the interface, and thus contributing to the growth of $\tilde{h}$, while $\alpha'(1+ \beta'\kappa(y))$ denotes the rate of particle escaping  with the effect of local interface curvature ($\kappa(y) = \pp^2 \tilde{h}/ \pp y^2$) taken into account. The noise terms are $b_i(x)$ which are Markovian, spatially independent and are either 0 or 1 with equal probability. Since the interface does not move at the steady state by assumption, $\alpha$ has to be identical to $\alpha'$. From now on, I will focus exclusively on the hydrodynamic limits (large temporal and spatial scales). So let us coarse grain $\tilde{h}$ by defining a new coarse-grained height function $h(y)$:
\beq
h(y) \equiv \frac{1}{\ell} \int_{y-\ell/2}^{y+\ell/2} \tilde{h}(y') \dd y'
\eeq
where $a \ll \ell \ll L_y$ and $\ell$ is large enough that $ \int_{y-\ell/2}^{y+\ell/2} b_i(y') \dd y'$ become Gaussian as a result of the central limit theorem. The EOM of $h(y)$ is then
\beqn
\frac{\pp h}{\pp t} &=& \alpha \left[g_1+\frac{1}{2} - (1+ \beta\kappa) \left(g_2+\frac{1}{2}\right)\right]
\\
\label{eq:hew}
&=& \alpha \left[\frac{\beta\kappa}{2} +\beta \kappa g_2 +(g_1-g_2)\right]
\eeqn
where $g_i$ are now Gaussian noises such that
\beqn
\la g_i(y,t) \ra&=&0
\\  \la g_i(y,t) g_j(y',t')\ra&=&\frac{a}{4\ell }\delta_{ij}\delta (t-t')\delta(y-y')
\ .
\eeqn
In the long-wavelength limit, the fluctuating term $\alpha \beta \kappa g_2 \sim \pp^2h/\pp y^2 \rightarrow 0$  and so the only relevant fluctuations come from the Guassian fluctuations $\alpha(g_1-g_2)$. Therefore, in the hydrodynamic limits, \eq (\ref{eq:hew}) is exactly the Edwards-Wilkinson model \cite{edwards_prsa82}, with the effective surface tension given by $\alpha \beta/2$. 
{ Note that the effective surface tension here is always positive, which is a requirement for having a stable interface. As such this effective surface tension is distinct from the mechanical definition of the surface tension, which has been shown to be negative in MIPS \cite{bialke_prl15}. As mentioned in Section 2.2, the negative tension around the interface is likely to be caused by the particles  escaping from the condensed phase that are now travelling close to being parallel to the interface. Consistent with our definition of the location of the interface (\fig \ref{fig:interfaces}), these escaped particles are not considered  part of the condensed phase  and are thus ignored in our discussion of the interfacial fluctuations. In other words, the effective surface tension derived here is distinct from the mechanical definition of the surface tension discussed by Bialke {\it et al} \cite{bialke_prl15}.
 }

Given that our stochastic model is equivalent to the Edwards-Wilkinson model, the temporal and steady-state dynamics of interfacial width is known to following the scaling form:
\beq
w_L(t) = L^\alpha \cF_{\rm EW} \left(\frac{t}{L^{\alpha/\beta}} \right)
\eeq
for some scaling function $\cF_{\rm EW}(.)$ (\fig \ref{fig:ew}). In a 2D system where the interface is a line, $\alpha=1/2$ and $\beta=1/4$ \cite{edwards_prsa82,barabasi_b95}.  As shown in \fig \ref{fig:ew}, these expectations are confirmed with direct simulation of a discretised version of the original EOM in \eq (\ref{eq:tildeh}). This model thus 
{ provides an analytical argument supporting}
the  steady state scaling $w_L(t \rightarrow \infty)\sim L^{1/2}$  recently observed numerically \cite{bialke_prl15}.

To summarise this section, I have incorporated the caging effect as discussed in Sect.~\ref{sec:curved} into the modelling of the stochastic dynamics of particle exchanges at the interface. The model equation is then shown to be equivalent to the Edward-Wilkinson model in the hydrodynamic limits. In particular, the emergence of the {\it effective surface tension} term ($\alpha\beta \kappa/2$) from the particle dynamics at the interface also explains why the interface is flat when both phases span the system, and circular when one phase does not span the system.

\section{MIPS in 3D}
I have so far analysed the interfacial properties in MIPS in 2D using a combination of analytical and numerical methods. Here I will extrapolate the results obtained to MIPS in 3D. 

\subsection{Flat interface}
Employing the sharp interface model for MIPS in 3D, the { pressure} balance equation in \eq (\ref{eq:forceb}) becomes
\beq
\frac{a\rho_c f_a}{2\pi} +p_r =\frac{\rho_v f_a^2}{6 \eta D_r}
\ ,
\eeq
where  on the R.H.S.\ the active force contribution corresponds to the swim pressure of active particles in 3D in the vapour phase \cite{takatori_prl14}, and on the L.H.S.,
the first terms comes from  active contribution to the force assuming again that the orientation is isotropic in the condensed phase. With regards to the minimal active force required for MIPS, using the approximations that $f_r$ is negligible and that $\rho_c/\rho_v \simeq 1$, we arrive at 
\beq
f_a^{min} \geq \frac{3a \eta D_r}{\pi}
\eeq
which is very similar to the expression in 2D (\eq (\ref{eq:minf})).

\subsection{Spherical interface}
The dimensional analysis presented in \Sec \ref{sec:gtrel} applies also spherical drops in 3D. Therefore, if the assumption that the escape range decreases with the curvature of the drop, then the Gibbs-Thomson relation should emerge in the large drop limit (\eq  (\ref{eq:rho_R})). In particular, we again expect the MIPS coarsening kinetics to be equivalent to the equilibrium scheme at low supersaturation \cite{lifshitz_jpcs61,yao_prb92,wittkowski_natcomm14}.

\subsection{Interface fluctuations}
For MIPS in 3D, the interface is two dimensional and so there are two principal curvatures. If we adopt the natural assumption that the escape range now depends on the mean curvature, then the theoretical analysis presented in \Sec \ref{sec:fluctuations} applies in 3D straightforwardly. As a result, keeping only the linear terms will again lead to the Edwards-Wilkinson model in the hydrodynamic limits.

\section{Discussion \& Outlook} 
In this paper I have investigated the microscopic dynamics of active particles in the interfacial regions in MIPS using a combination of simulations and analytical arguments, and   demonstrated
i) how interface stability is achieved, ii) why the  GT relation emerges in MIPS, and iii) how interface fluctuations scale with the system size. Therefore, I have shown that all the observed ``surface tension'' related phenomena found in MIPS result from the  microscopic dynamics of the active particles. More specifically, I have demonstrated that { pressure} balance in MIPS is achieved  because of the orientation anisotropy in the region, which leads to a high active force directed towards the condensed phase. By incorporating the caging effects of neighbouring particles in the peripheral of a condensed drop, I have shown how the Gibbs-Thomson relation emerges naturally in MIPS, which dictates that the larger the condensed drop, the smaller the vapour concentration outside the drop. If the supersaturation level is small, the GT relation leads to diffusive transfer of active particles from small drops to larger drops. As a result, the late-stage coarsening kinetics in an active phase-separating system should follow the temporal scaling as in equilibrium phase separation: i.e., the average droplet size in the system $\la R(t) \ra$ goes like  $t^{1/3}$ \cite{lifshitz_jpcs61,yao_prb92,wittkowski_natcomm14}. In addition, the droplet size distribution  should approach asymptotically the universal size distribution obtained by Lifshitz and Slyozov \cite{lifshitz_jpcs61}.
Lastly, motivated by the same caging effects, I have proposed a stochastic model that describes the interfacial fluctuations in MIPS. In this model, the probability  of particles leaving the interface is assumed to be dependent on the interface curvature. Analytical argument is then provided to show that the proposed model belongs to the same universality class of the Edwards-Wilkinson model.

There are a number of future directions that are of interest. For instance,
phase separation may play a role in the cytoplasmic re-organisation during asymmetric cell division \cite{brangwynne_science09,lee_prl13}. How the activity in the cytoplasm due to the many motor proteins contribute to such re-organisation via phase separation awaits more attention.
Moreover, the fact that active phase separation occurs naturally begs the question of the existence of the critical point as in the equilibrium case.  The critical transition in incompressible active fluids has recently been shown to give rise to a novel universality class \cite{chen_njp15}.
And if a critical transition  exists in MIPS,  will the critical exponents be identical to those in the equilibrium case which belong to the 2D Ising Universality class? This question awaits further investigation.

\section*{Appendix}
\appendix
\section{Simulation details}
\subsection{Particle dynamics simulation}
\label{sec:a1}
For the simulation results reported in \Sec \ref{sec:flat} (Figs~\ref{fig:mainpic} \& \ref{fig:interfaces}), I numerically integrating the Langevin equations for each particle in the bulk of the system of size $L_x \times L_y$ by using the following updates:
\beqn
\label{eq:update1}
\theta_i^{t+\tri t} &=&\theta_i^t   + \sqrt{2D_r \tri t} g_i^t
\\
\label{eq:update2}
r_i^{t+\tri t} &=&-\frac{1}{\eta} \sum_{j\neq i}\nabla_{\bbr_i} U(|\bbr^{t}_i-\bbr^{t}_j|) +\frac{f_a}{\eta}\cos \theta^t_i
\ ,
\eeqn
where $g_i^t$ are Gaussian distributed random variables with zero mean and unit variance, $U$ is given by the  Weeks-Chandler-Andersen potential shown in \eq (\ref{eq:wca}). Periodic boundary condition is enforced in the $y$, direction; while in the $x$ direction, if { $x_i^{t+\tri t}<0$, then it is reset to zero, and if $x_i^{t+\tri t}>L_x$, then $x_i^{t+\tri t}=L_x$} and $\theta_i^{t+\tri t}$ is an angle chosen at random.

The parameters of the simulations are: $\tri t =10^{-5}$, $a=1$, $\eta=1$, $f_a=100$, $D_r=3$, $A=25/6$ for repulsive particles and $A=0$ for point particles. The system has width $L_x=50$ and height $L_y=10\sin(\pi/3)$, with 300 particles initialized in the configuration of a hexagonal lattice (with spacing 1) next to the left boundary, and  random orientations. Two hundred million time steps are evolved to equilibrate the system and then data are collected in the subsequent two hundred million time steps.

\subsection{Point particles in an annular geometry}
\label{sec:a2}
For the simulation results reported in 
\Sec \ref{sec:curved} (Figs~\ref{fig:Cinterface}), the system is now annular with inner radius $R$ and outer radius $R+L_r$. The same updates as in \eqs (\ref{eq:update1}) and (\ref{eq:update2}) are performed for the point particles in the bulk of the system. Concerning the boundary conditions,
if $r^t\leq R$ then the particle becomes part of the condensed drop periphery, and the orientation follows the update:
\beqn
\theta_i^{t+\tri t} &=&\theta_i^t + \sqrt{2D_r \tri t} g_i^t
\ ,
\eeqn
while the position remains the same
until $\psi^t\equiv |\theta^t-\varphi^t| < \sigma_R$ (\fig \ref{fig:Cinterface}(b)), in which case the particle leaves the condensed phase. If $r_i^{t+\tri t}>R+L_r$, then $r_i^{t+\tri t}=R+L_r$ and $\theta_i^{t+\tri t}$ is an angle chosen at random.

{
For distinct annulus geometry,  the density $\rho_R(r)$ is normalised by  $\rho_c$, which is the density of particles at the inner circular wall. This is based on the assumption that $\rho_c$ is the same irrespective of drop sizes.
}

The parameters of the simulations are: $\tri t =10^{-3}$,  $\eta=1$, $f_a=100$, $D_r=3$, $A=0$. The system has 1000 particles initialized with random orientations and positions.  Twenty million time steps are evolved to equilibrate the system and then data are collected in the subsequent twenty million time steps. Simulations are performed for $R=20,40,60,80$ and $100$ while $L_r$ is always  $50$.

\subsection{Fluctuating interfaces}
\label{sec:a3}
To simulate interface fluctuations according to \eq (\ref{eq:tildeh}), I discretise the interface (a line) into $L$ sites with height values $h_i$ where $i=1,\ldots, L$. Periodic boundary condition is enforced. The updates are performed as follows:
\beqn
\nonumber
h_i^{t+\tri t} &=&h_i^t +[1+\beta(h_{i-1}^t+h_{i+1}^t-2h_i^t)]\tilde{g}_i^t
\\
&&-[1-\beta(h_{i-1}^t+h_{i+1}^t-2h_i^t)]\tilde{g}_i^t
\ ,
\eeqn
where $\tilde{g}_i^t$ are either 0 or 1 chosen with equal probability, and $\beta=0.1$. The system sizes simulated are $L=160,200,240,280,320$ and $360$. The total number of time steps simulated for each system size is $5L^2/2$.

\vspace{2cm}
I thank Christoph Weber (Max Planck Institute for the Physics of Complex Systems),
Fernando Peruani (Universit\'{e} Nice Sophia Antipolis) and Richard Sear (University
of Surrey) for stimulating discussions
and for their comments on the manuscript.


\end{document}